# Synergetic capacity planning of private and public EV charging piles via city-scale multiobjective optimization

Yiwu Hao [1, †], Hong Yuan [1, †], Nan Zhou [2], Minda Ma [3 *]

1. School of Management Science and Real Estate, Chongqing University, Chongqing, 400045, P. R. China
2. Energy and Resources Group, University of California, Berkeley, CA 94720, United States
3. School of Architecture and Urban Planning, Chongqing University, Chongqing, 400045, P. R. China

- Corresponding author: Prof. Dr. Minda Ma, Email: minda.ma@cqu.edu.cn;
  Homepage: https://globe2060.org/MindaMa/

## Highlights

- Proposed a high-resolution EV charging capacity planning model for public and private piles.
- EV electricity use in Chongqing tripled, reaching 57.5 GWh between mid-2022 and the end of 2024.
- Charging piles were concentrated in the urban core, while the public charging supply lagged demand.
- The optimized charging capacity allocation achieved better balance, reducing the score from 0.65 to 0.28.
- By 2030, Chongqing was projected to require 1.8 million charging piles with a 9:1 private-to-public ratio.

## Abstract

Rapid electric vehicle (EV) expansion necessitates optimized charging infrastructure to bridge the persistent gaps between vehicle growth and charger availability. This study develops a demand-driven framework for city-scale EV charging demand assessment and charging pile capacity planning. It employs a bottom-up estimation approach to quantify electricity demand and a Harris Hawks Optimization algorithm to solve capacity planning challenges, capturing spatiotemporal demand variations across powertrain types and guiding allocation over 2022–2030 in Chongqing, China. The results show that (1) compared with June 2022, monthly EV electricity consumption tripled to 57.5 gigawatt-hours by the end of 2024, characterized by significant seasonal volatility and a structural shift in which the combined share of plug-in hybrid electric vehicles and extended-range electric vehicles reached 57.6%, necessitating a transition toward technology-specific infrastructure planning; (2) historical evaluations reveal a marked spatial mismatch, with actual deployment heavily concentrated in the urban core while public charging capacity consistently lagging behind demand, whereas the proposed optimized configuration achieved a superior comprehensive performance score of 0.28, compared to 0.65 for actual deployment, in balancing service adequacy across the “Core-Suburban-Exurban” hierarchy; and (3) by 2030, Chongqing is projected to require approximately 1.8 million charging units to sustain a stable 9:1 private-to-public ratio, a synergetic strategy expects to significantly mitigate urban–rural service disparities and enhance overall system resilience and grid compatibility. Ultimately, this study provides a versatile, spatially explicit tool for policymakers to support sustainable and cost-effective EV infrastructure deployment aligned with long-term electrification targets.

## Abbreviations and symbols

### Abbreviation notation

BEV – Battery electric vehicle

CLTC – China Light-Duty Vehicle Test Cycle

EV – Electric vehicle

EREV – Extended-range electric vehicle

FGM – Fractional-order Grey Model

GWh – Gigawatt-hours

HHO – Harris Hawks Optimization

NEDC – New European Driving Cycle

WLTP – Worldwide Harmonized Light Vehicles Test Procedure

PHEV – Plug-in hybrid electric vehicle

### Nomenclature

$AvgLoad$ – Annual average grid load

$Capacity_i$ – Battery capacity of vehicle version $i$

$CLTC_i$ – Rated driving range of vehicle version $i$ under the CLTC cycle

$C_{\text{pub},i}$ – Number of public charging piles in district $i$

$C_{\text{pri},i}$ – Number of private charging piles in district $i$

$C_{\text{cap,pub},i}$ – Unit investment cost of public charging piles in district $i$

$C_{\text{cap,pri},i}$ – Unit investment cost of private charging piles in district $i$

$C_{\text{om,pub},i}$ – Annual operation and maintenance costs of public charging piles in district $i$

$C_{\text{om,pri},i}$ – Annual operation and maintenance costs of private charging piles in district $i$

$CRF$ – Capital recovery factor for annualization

$E_k$ – Annual total EV charging demand in city $k$

$K_{pub}$ – Annual equivalent utilization rate of one public pile

$K_{pri}$ – Annual equivalent utilization rate of one private pile

$L_s^{base}$ – Base grid load in quarter $s$ excluding charging demand

$Load_s$ – Total grid load in quarter $s$

$Mileage_k$ – Average annual driving mileage in city $k$

$N$ – Number of administrative districts in city $k$

$n$ – Project life in years

$p_{pub,i,s}$ – Average charging power of one public pile in district $i$ during quarter $s$

$p_{pri,i,s}$ – Average charging power of one private pile in district $i$ during quarter $s$

$r$ – Discount rate

$Registrations_j$ – Registered vehicles of model $j$

$v_i$ – Charging demand of vehicle version $i$ in the extreme-climate season

$Z_1$ – Total life-cycle cost

$Z_2$ – Variance of quarterly grid load

$Z_3$ – Total energy supply potential

$Z'_k$ – Normalized value of objective $k$

$Z_{k,min}$ – Minimum value of objective $k$

$Z_{k,max}$ – Maximum value of objective $k$

$\gamma_i$ – Registration share of vehicle version $i$ within model $j$

$\lambda_i$ – Energy-consumption correction factor in the mild season

$\rho_i$ – Energy-consumption correction factor in the extreme-climate season

$\Theta_j$ – Annual charging demand of all versions under model $j$

$\Omega_i$ – Annual charging demand of vehicle version $i$

$\omega_i$ – Charging demand of vehicle version $i$ in the mild season

# 1. Introduction

## 1.1. Background

The global shift toward low-carbon transport is accelerating in response to climate goals, air-quality pressures, and energy-security concerns, making electric vehicles (EVs) a key part of decarbonizing mobility [1, 2]. This rapid uptake has pushed charging demand upward and brought charging pile capacity configuration to the forefront of infrastructure planning, where the goal is to align charging capacity with demand across space and time while avoiding both shortages and underused investments [3, 4]. China, as the world's largest EV market, reached 31.4 million EVs and more than 12 million charging piles by 2024, yet the growth of vehicles and charging infrastructure has not been fully synchronized [5, 6]: the national vehicle-to-pile ratio has remained on a multiyear plateau since approximately 2020, fluctuating roughly between 2.5:1 and 3:1, and spatial imbalances persist across regions and within cities [7]. Chongqing, a national pilot city and major industrial hub in western China, has issued successive plans to expand and upgrade its charging network, including recent initiatives centered on fast and ultra-fast charging ; however, strong differences in development and travel patterns across districts and counties continue to produce uneven capacity distribution between core urban areas and peripheral zones, highlighting the need for demand-oriented, spatially explicit capacity configuration tailored to Chongqing's local context.

## 1.2. Literature review

The accurate estimation of EV charging demand is fundamental for effective infrastructure planning, ensuring system reliability, operational efficiency, and equitable access [8, 9]. Early studies often relied on simplified projections that aggregated all EVs or only partially distinguished between powertrain types, potentially distorting city-scale demand estimates and hindering the fair distribution of charging resources [10, 11]. Recent studies have emphasized differentiating among battery electric vehicles (BEVs), plug-in hybrid electric vehicles (PHEVs), and extended-range electric vehicles (EREVs), acknowledging their distinct energy consumption profiles and usage patterns [12]. Approaches to demand estimation include data-driven models, for example, deep learning [13, 14], large language models [15, 16], and time-series regressions [17, 18], which capture spatiotemporal patterns but require extensive historical data and face challenges in generalization; simulation models, which integrate traffic flows [19], grid interactions [20, 21], and behavioral assumptions [22, 23] to provide macrolevel insights; and bottom-up models [24], which aggregate individual vehicle charging behaviors and thereby offer greater accuracy in heterogeneous urban contexts.

Charging infrastructure planning has been widely studied using location–allocation [25, 26], network flow [27, 28], and optimization-based [29] approaches. Microlevel studies have focused on optimizing individual station siting, sizing, and operational schedules, often integrating grid-side constraints such as transformer capacity and voltage limits [30]. Multi-objective frameworks have been developed to balance economic cost [31], utilization efficiency [32], service quality [33, 34], and spatial equity [35]. Some studies extended to highway or intercity networks using flow-refueling models that simulate vehicle routing and charging behavior [36, 37]. However, most studies remain limited to station-level or subsystem analysis [38], with insufficient capacity to address city-wide trade-offs among cost, grid stability, and spatial equity under long-term electrification targets [39].

A variety of optimization algorithms have been employed to solve these planning problems [40, 41]. Traditional optimization approaches, including mixed-integer linear programming [42] and dynamic programming [43], generally handle small or fully deterministic problems effectively, but they encounter difficulties when applied to the high-dimensional, nonlinear, and uncertain nature of urban EV charging networks. To overcome these limitations, many studies increasingly relied on metaheuristic and bioinspired algorithms. Techniques such as ant colony optimization [44], grey wolf optimization [45], differential evolution [46], and the Harris Hawks Optimizer (HHO) [47] offer the ability to explore large solution spaces, adapt to multiple objectives, and maintain robustness under uncertainty. In addition, advanced or hybrid versions of these algorithms speed up convergence and generate more solution options, making city-scale planning feasible while handling complex operational and spatial constraints.

Recent studies have highlighted the importance of integrating temporal and spatial heterogeneity, demand uncertainty, and user behavior into capacity planning models [48, 49]. Various methods, including stochastic simulations, scenario analysis, and machine learning-based load forecasting, have been employed to dynamically allocate charging piles. Moreover, increasing attention to equity has led to the incorporation of fairness constraints that are based on population and mobility [50], ensuring that charging infrastructure benefits are not concentrated in high-demand districts but distributed more evenly across the city. Collectively, these studies suggest that robust planning requires models that simultaneously address demand fluctuations, uncertainty, and fairness, highlighting a shift from purely efficiency-oriented approaches toward more holistic, city-scale strategies.

### *1.3. Motivation, contributions, and the organization*

However, several gaps continue to hinder effective charging pile capacity configuration under large-scale electrification. First, demand inputs used for planning are often derived from aggregate, stock-based projections, which cannot fully capture differences across powertrain types, market composition, and seasonal climate effects and therefore provide limited support for high-resolution city planning [51]. Second, city-scale planning remains fragmented: public and private charging piles are frequently planned without a unified allocation logic across all districts [52], and regional equity is often treated as a general principle rather than implemented as an enforceable planning requirement [53]. Third, existing studies have paid less attention to nonlinear growth patterns in infrastructure deployment, which weakens forward-looking capacity trajectories that underpin medium- to long-term plans. Addressing these limitations is essential for developing scalable, practical planning strategies that balance cost, grid stability, and equitable access. Three key questions thus arise from these gaps and warrant further investigation:

- How can EV charging demand be estimated across powertrain types?
- How can multi-objective modeling optimize Chongqing's charging for cost, grid, and equity?
- How should Chongqing optimize public and private charging deployment through 2030?

To address the identified challenges, this study develops a planning framework for citywide charging pile capacity configuration that is driven by energy demand. First, it proposes a bottom-up model using data from 2022 to 2024 to quantify charging demand dynamics and structural differences across powertrain types at seasonal-to-monthly resolution. Second, it introduces a nonlinear forecasting component to characterize the evolution of demand and charging piles capacity over time up to 2030, providing consistent inputs for medium-term planning. Third, it formulates a city-scale optimization model with multiple objectives that co-plans public and private charging piles and embeds regional equity as a rigid constraint so that the resulting allocation balances cost efficiency, grid compatibility, and fair access. Chongqing is selected as the case study given its rapid electrification, strong policy push, and pronounced spatial heterogeneity, which together make it a representative setting for testing citywide capacity configuration methods.

**The principal contribution of this study** lies in its development of an integrated, city-scale planning framework for EV charging piles that integrates high-resolution demand accounting, forward-looking capacity evolution, and allocation based on multiple objectives. By explicitly representing technology, market, and climate drivers, the demand estimation component provides more credible inputs than conventional aggregate models. The planning framework further incorporates economic, grid, and equity considerations, enabling transparent trade-off analysis and more balanced charging pile deployment. This study addresses a critical gap in existing studies, which often focus on station-level optimization or rely on simplified demand assumptions, limiting applicability to citywide, multiyear planning. While Chongqing serves as a demonstrative case, the methodology is designed to be transferable to other cities experiencing rapid electrification and spatial heterogeneity, offering a scalable tool to guide long-term investment, policy design, and scenario evaluation.

The remainder of this paper is organized as follows: Section 2 describes the proposed framework, including the demand accounting model, the multi-objective planning formulation, and the data sources. Section 3 analyzes Chongqing's historical EV charging demand and piles, evaluates actual versus optimized capacity configurations, and validates the planning framework. Section 4 presents projected charging pile growth from 2025 to 2030, the optimized spatial allocation across districts, and related policy implications. Section 5 concludes the paper and outlines directions for future work.

## 2. Materials and methods

This section presents an integrated framework for EV charging demand estimation and city-scale infrastructure planning. Section 2.1 develops a bottom-up demand model for BEVs, PHEVs, and EREVs; Section 2.2 formulates a multi-objective capacity planning model solved by HHO; and Section 2.3 summarizes the data sources and parameter settings for the Chongqing case study.

### *2.1. High-resolution bottom-up model for charging demand*

Considering that EV energy consumption is affected by temperature, particularly in summer and winter due to increased cooling and heating loads from air conditioning [54], charging demand exhibits clear seasonal variation. Accordingly, the year is divided into a mild season (spring and autumn) and an extreme-climate season (summer and winter), and separate charging demand models are established for each vehicle version. For a given version $i$ in city $k$, the annual charging demand is calculated as follows:

$$\omega_i = \gamma_i \cdot Registrations_j \cdot Mileage_k \cdot \frac{Capacity_i}{\lambda_i \cdot CLTC_i} \tag{1}$$

$$\upsilon_i = \gamma_i \cdot Registrations_j \cdot Mileage_k \cdot \frac{Capacity_i}{\rho_i \cdot CLTC_i} \tag{2}$$

where $\omega_i$ and $\upsilon_i$ (kilowatt-hours, kWh) denote the charging demand of version $i$ in the mild season and extreme-climate season, respectively. $\gamma_i$ is the registration share of version $i$ within model $j$; $Registrations_j$ is the number of registered vehicles of model $j$; $Mileage_k$ (kilometers, km) is the average annual driving mileage of private cars in city $k$; $Capacity_i$ (kWh) is the battery capacity of version $i$; $CLTC_i$ (km) is the rated driving range of version $i$ under the China Light-Duty Vehicle Test Cycle (CLTC); and $\lambda_i$ and $\rho_i$ are the actual energy-consumption correction factors for the mild season and the extreme-climate season, respectively.

Based on the seasonal results, the annual charging demand of version $i$ is obtained by aggregation:

$$\Omega_i = \omega_i + \upsilon_i \tag{3}$$

Aggregating the demand of all versions under model $j$ yields the annual total demand of that model:

$$\Theta_j = \sum_{i=1}^{l} \Omega_i \tag{4}$$

Summing the demand of all models in city $k$ gives the city's annual total electric vehicle charging demand:

$$E_k = \sum_{j=1}^{m} \Theta_j \tag{5}$$

This can be further rewritten as:

$$E_k = \sum_{j=1}^{m} \sum_{i=1}^{l} \gamma_i \cdot Registrations_j \cdot (Mileage_k \cdot \frac{Capacity_i}{\lambda_i \cdot CLTC_i} + Mileage_k \cdot \frac{Capacity_i}{\rho_i \cdot CLTC_i}) \tag{6}$$

*2.2. Synergetic capacity planning model for charging piles*

The charging pile capacity planning model is designed to minimize the total life-cycle cost, minimize grid load fluctuation, and maximize the energy supply potential. The first objective is to minimize the total life-cycle cost of the charging piles:

$$min\, Z_1 = \sum_{i=1}^{N} (C_{cap,pub,i} \cdot CRF + C_{om,pub,i} \cdot C_{pub,i} + C_{cap,pri,i} \cdot CRF + C_{om,pri,i} \cdot C_{pri,i}) \tag{7}$$

where $Z_1$ is the total life-cycle cost. $N$ is the number of administrative districts in city $k$. $C_{\text{pub},i}$ and $C_{\text{pri},i}$ are the numbers of public and private charging piles allocated to district $i$, respectively. $C_{\text{cap,pub},i}$ and $C_{\text{cap,pri},i}$ are the unit initial investment costs of public and private charging piles in district $i$, respectively. $C_{\text{om,pub},i}$ and $C_{\text{om,pri},i}$ are the corresponding annual operation and maintenance costs. $CRF$ is the capital recovery factor, used to convert the initial investment into an annualized equivalent. The capital recovery factor is defined as:

$$CRF = \frac{r(1+r)^n}{(1+r)^n - 1} \tag{8}$$

where $r$ is the discount rate and $n$ is the project life in years.

The second objective is to reduce grid load fluctuation:

$$min\, Z_2 = \frac{1}{4} \sum_{s=1}^{4} (Load_s - AvgLoad)^2 \tag{9}$$

where $Z_2$ is the variance of quarterly grid load. $Load_s$ is the total grid load in quarter $s$, and $AvgLoad$ is the annual average load. The index $s$ runs from 1 to 4, representing the four quarters of the year. The quarterly load is calculated as:

$$Load_s = \sum_{i=1}^{N} (C_{pub,i} \cdot p_{pub,i,s} + C_{pri,i} \cdot p_{pri,i,s}) + L_s^{base} \tag{10}$$

where $p_{pub,i,s}$ and $p_{pri,i,s}$ are the average charging power of one public pile and one private pile in district $i$ during quarter $s$. $L_s^{base}$ is the base grid load in quarter $s$, excluding charging demand. The annual average load is given by:

$$AvgLoad = \frac{1}{4}\sum_{s=1}^{4} Load_s \tag{11}$$

where $AvgLoad$ is simply the mean of the four quarterly loads.

The third objective is to maximize the energy supply potential of the charging network:

$$max\ Z_3 = \sum_{i=1}^{N} (C_{pub,i} \cdot K_{pub} + C_{pri,i} \cdot K_{pri}) \tag{12}$$

where $Z_3$ is the total energy supply potential. $K_{pub}$ and $K_{pri}$ are the average annual equivalent utilization rates of one public pile and one private pile, respectively.

Because the three objectives have different scales and meanings, they are first normalized before being combined into a single objective. For the two minimization objectives, the normalized form is:

$$Z'_k = \frac{Z_k - Z_{k,min}}{Z_{k,max} - Z_{k,min}}, k = 1,2 \tag{13}$$

where $Z'_k$ is the normalized value of objective $k$. $Z_{k,min}$ and $Z_{k,max}$ are the minimum and maximum values of objective $k$ across the search space.

For the maximization objective, the equivalent normalized cost form is:

$$Z'_3 = \frac{Z_{3,max} - Z_3}{Z_{3,max} - Z_{3,min}} \tag{14}$$

After normalization, the three objectives are combined into one weighted objective function:

$$min\ Z = w_1 Z'_1 + w_2 Z'_2 + w_3 Z'_3 \tag{15}$$

where $Z$ is the final composite objective. $w_1$, $w_2$, and $w_3$ are the normalized weights of the three objectives, with $w_1 + w_2 + w_3 = 1$ and $w_i \geq 0$.

The feasibility of the optimization result is ensured through a set of composite constraints, as detailed in Appendix B, together with the HHO-based solution procedure.

### *2.3. Data sources*

This study focused on Chongqing, one of China's fastest-growing EV clusters, supported by a strong industrial base, policy incentives, and a diverse urban form. The study examined EV models registered in Chongqing from June 2022 to December 2024, including BEVs, PHEVs, and EREVs, to assess operational charging demand and determine a public–private charging pile capacity plan. EV registration data were obtained from Chezhuzhijia (https://www.16888.com/), which provides regularly updated vehicle registration statistics in China. Technical specifications, including battery capacity and CLTC range, were collected from Autohome (https://www.autohome.com.cn/), which compiles specification information from manufacturers and official sources. Estimated annual mileage in Chongqing was taken from Ou et al. [55], providing localized behavioral evidence. Existing charging pile stock data in Chongqing were compiled from the China Electric Vehicle Charging Infrastructure Promotion Alliance database (https://evcipa.com/index). The detailed sources and settings of the model input parameters are provided in Appendix C. By integrating these sources, this study established a transparent and traceable data foundation, improving the credibility and practical value of the planning scheme.

## 3. Results

Building on Section 2, this section presents three findings from the Chongqing case: the spatiotemporal evolution of charging demand for BEVs, EREVs, and PHEVs, the superiority of the optimized charging piles capacity configuration over the actual deployment, and the robustness and validity of the planning framework confirmed through validation and repeated runs.

### *3.1. Spatiotemporal patterns of charging demand*

This study adopted a bottom-up, vehicle-model-level approach to estimate the energy intensity of BEVs, EREVs, and PHEVs in Chongqing from June 2022 to December 2024. By examining the relationship between energy intensity and registration volume, the analysis identified technology efficiency, consumer preference, and competition patterns and provided inputs for charging pile capacity planning. Fig. 1 summarizes the relationship between energy intensity and registration volume for EV models. It compared BEV, EREV, and PHEV models to identify differences in energy intensity, consumer preference, and market competition. As shown in Fig. 1 A, BEVs such as Model Y, Yuan PLUS, and AION S combined high registration volumes with relatively low energy intensity, generally in the range of 11.8–13.3 kWh/100 km, indicating a strong link between technical maturity and market acceptance. Micro EVs such as the Hongguang MINIEV and Changan Lumin achieved even lower consumption, at approximately 8.4–8.9 kWh/100 km, but they served more specialized urban use cases. In the EREV market, models such as Deepal SL03 and Deepal S7 showed both relatively low energy intensity and strong market penetration, while premium models including the Li L series and AITO M9 had higher consumption but remained competitive because of their positioning for family use and features focused on comfort. The PHEV market was more heterogeneous: BYD Song PLUS and Qin PLUS performed well in both sales and energy intensity, whereas larger models such as the Tank 500 New Energy and Wey Lanshan-PHEV had higher energy intensity but retained demand in specific market segments.

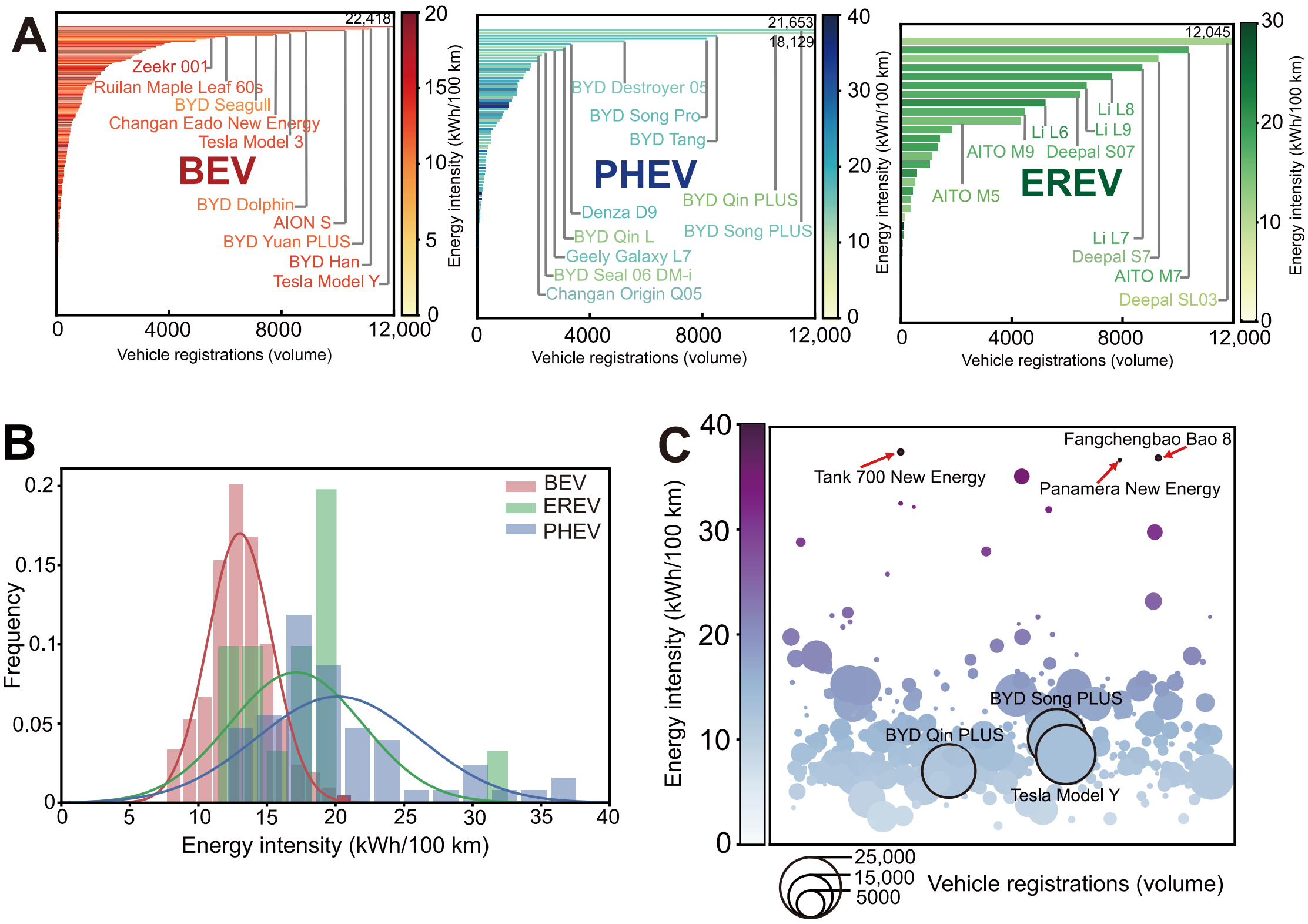

**Fig. 1.** Distributions of EV registrations and energy intensity in Chongqing, June 2022 to December 2024. (A) Vehicle registration and energy intensity patterns for BEVs, EREVs, and PHEVs; (B) frequency distribution of energy intensity by EV type; (C) integrated analysis of energy intensity and registrations across all EV types.

Fig. 1 B shows distinct energy intensities across EV types. BEVs exhibited the most concentrated distribution, peaking at 11–16 kWh/100 km, indicating relatively low variation and high energy efficiency. EREVs shifted to a slightly higher and wider range, peaking at 12–20 kWh/100 km due to the influence of range extension systems and energy management strategies. PHEVs showed the widest dispersion, with a peak at approximately 20 kWh/100 km and a long tail extending to 25–35 kWh/100 km. This broader spread suggested greater sensitivity to charging behavior and driving conditions.

Fig. 1 C indicates a close link between market scale and energy intensity in the EV market. Mainstream models such as Model Y, Song PLUS, and Yuan PLUS combined high registration volumes with low energy intensity, forming the core of the market. This pattern suggests that consumer demand depends not only on technical maturity but also on energy intensity, supported by manufacturers' advantages in battery systems, powertrain optimization, and energy management. The market can be divided into three clusters: high-volume, high-

efficiency models representing the mainstream path; low-volume, high-efficiency models relying on differentiation; and high-intensity models, including luxury and off-road vehicles, whose performance depended more on brand value and functionality. Notably, Tank 700 New Energy and Panamera New Energy had very high energy intensity but still attracted specific consumer groups. This suggested that in the premium EV segment, performance and brand identity outweigh energy intensity in purchasing decisions. Overall, the three EV types clearly differed in both energy intensity and market acceptance. These results support charging pile capacity planning and policy design.

Fig. 2 illustrates EV electricity consumption patterns in Chongqing from June 2022 to December 2024. The figure shows a clear upward trend in EV electricity demand across powertrain types and models. Fig. 2 A summarizes monthly and quarterly electricity consumption by EV type. Total monthly EV electricity consumption rose from 18.9 to 57.5 gigawatt-hours (GWh), an increase of 204%, indicating rapid market penetration [56]. Growth was uneven, remaining modest in late 2022 but accelerating in 2023 and early 2024. Seasonal variation was pronounced, with winter months generally higher than summer months. This pattern reflects colder weather, which reduces battery efficiency and increases heating demand, as well as more demanding operating conditions.

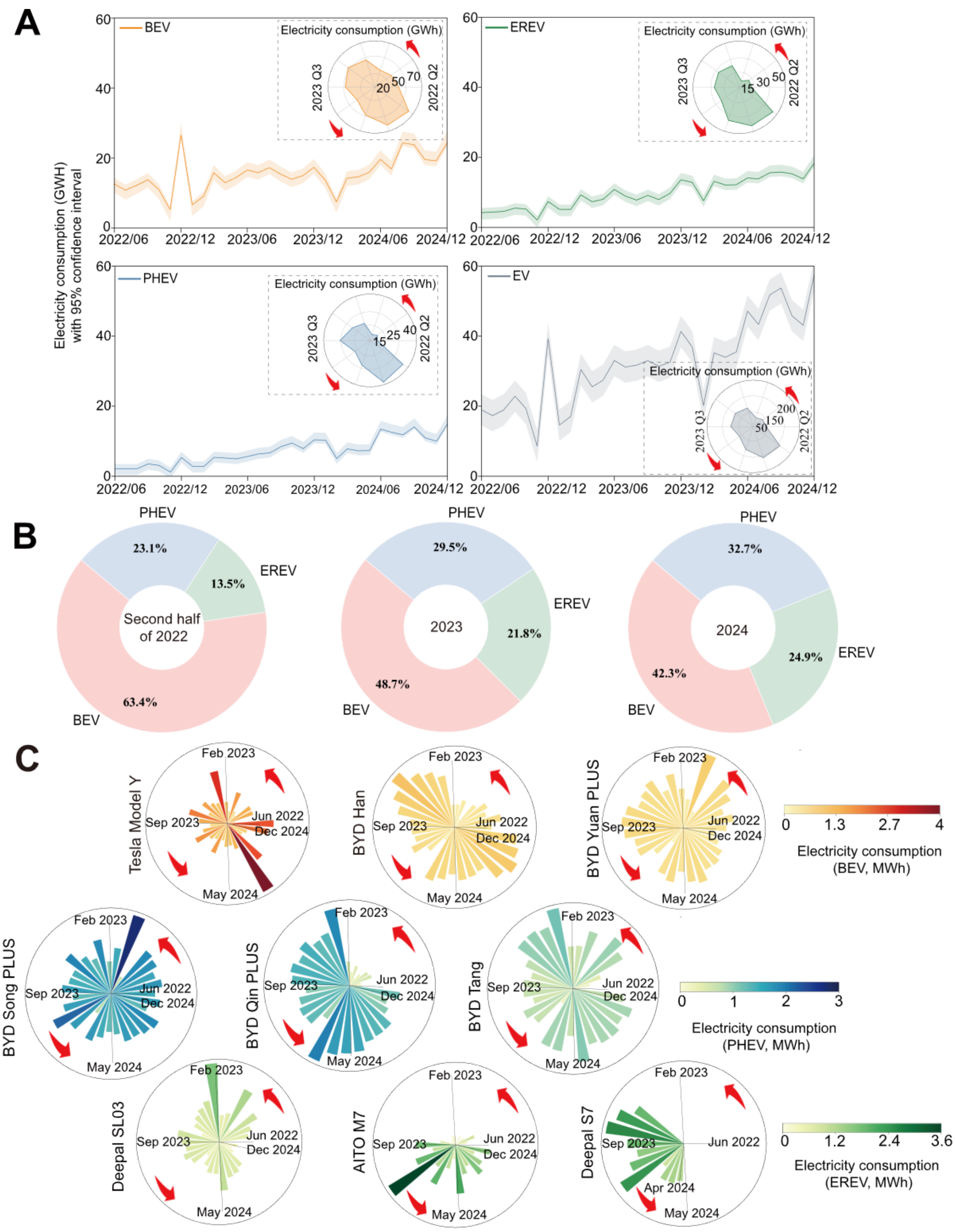

**Fig. 2.** Electricity consumption trends of different EV types in Chongqing, June 2022 to December 2024. (A) Time series trends in electricity consumption by EV type; (B) annual share of electricity consumption by EV type; (C) time series trends in electricity consumption for the top three BEV, PHEV, and EREV models by registrations.

Fig. 2 B shows the annual share of electricity consumption by powertrain type. BEVs remained the dominant contributor, despite their share declining from 63.4% in the second half

of 2022 to 42.3% in 2024. In contrast, PHEVs increased from 23.1% to 32.7%, while EREVs rose from 13.5% to 24.9%. Together, PHEVs and EREVs increased from 36.6% to 57.6%. This shift suggests a more diversified regional EV market and highlights the need for charging pile planning to account for differing demand characteristics across powertrain types.

Fig. 2 C highlights representative BEV, EREV, and PHEV models. Among BEVs, Tesla Model Y exhibited the highest electricity consumption and the largest monthly variation, with a range of approximately 8.3 times between its maximum and minimum values. BYD Han and Yuan PLUS showed more stable consumption and smoother growth. In the EREV group, Deepal SL03 and S7 showed clear seasonality, with higher consumption in winter and summer and lower consumption in spring and autumn. For PHEVs, BYD Song PLUS, Qin PLUS, and Tang accounted for most consumption. Song PLUS showed the strongest growth, Qin PLUS remained stable, and Tang presented more noticeable peaks around family travel periods. In summary, EV electricity demand is shaped by technology choice, model composition, user behavior, and external conditions rather than vehicle ownership alone.

Fig. D1 and its corresponding analysis are provided in Appendix D, showing that EV electricity demand in Chongqing was projected to rise continuously from 2025 to 2030, led by BEVs. The demand profile also exhibited clear seasonal variation, with higher values in summer and autumn. BEVs remain the principal driver of future electricity demand growth.

Overall, the above results reveal differences in energy intensity, market structure, and charging demand across BEVs, EREVs, and PHEVs in Chongqing from 2022 to 2024. These findings answer Question 1 raised in Section 1 and provide a solid basis for charging pile capacity planning.

### *3.2. Comparison of actual and planned pile capacity configurations*

From 2022 to 2024, Chongqing experienced rapid charging pile expansion. The gap between actual and planned deployment reflected early concentration in core demand areas and weak regional balance. By comparing the spatial distribution and coverage gradient, this study clarified the existing network and demonstrated the value of the planned configuration for resource optimization and long-term adaptability. Fig. 3 compares actual charging pile deployment in Chongqing with the proposed configuration from 2022 to 2024. It shows the spatial distribution of capacity placement and the performance differences between the two configurations. Fig. 3 A presents the spatial allocation of charging pile capacity. This result indicates that the actual configuration was highly concentrated in the urban core, with a much higher capacity density than in the surrounding suburban and county-level areas. This pattern matched stronger central demand but left peripheral areas with weaker coverage and accessibility. In contrast, the planned configuration retained a strong core focus while extending capacity more evenly to suburban districts and county centers, creating a more balanced spatial structure.

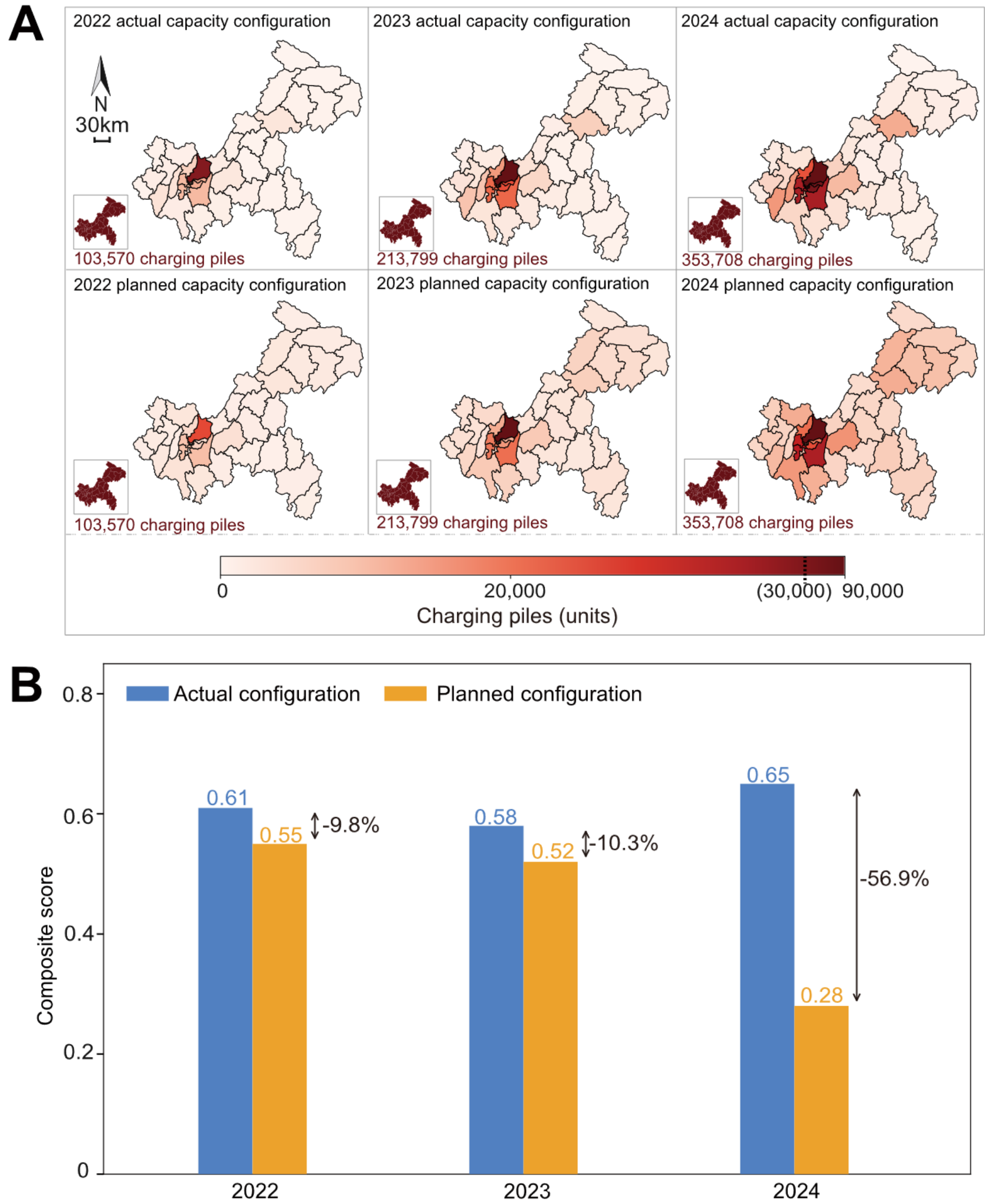

**Fig. 3.** Comparison of actual and planned capacity configurations of EV charging piles in Chongqing from 2022 to 2024. (A) Actual versus planned capacity configurations of EV charging piles; (B) comparative composite scores of actual and planned capacity configurations.

Fig. 3 B reports the comprehensive score and further confirms the advantage of the planned capacity configuration. Across all three years, the planned capacity configuration outperformed the actual deployment, and the gap became more pronounced over time. In 2024, the planned capacity configuration achieved a comprehensive score of 0.28, compared with 0.65 for the actual scheme, showing a clear improvement in overall system performance. This result suggests that the model did not merely chase a single target but instead achieved a more effective balance among cost control, grid stability, and service provision as demand expanded rapidly.

Overall, the comparison shows that the actual configuration met short-term urban-core demand but had weaknesses in regional balance and long-term adaptability. In contrast, the planned configuration preserved core-area efficiency while improving coverage in outlying areas and overall performance. These findings indicate that the proposed model can support charging pile planning that responds to current demand and aligns better with future growth, providing a more reliable basis for coordinated urban charging network development.

Fig. 4 compares actual and planned charging pile capacity configurations for private and public use across Chongqing's districts and counties in 2022, 2023, and 2024. Overall, the actual deployment showed a clear mismatch with the planning capacity configuration, while the planned configuration was more closely aligned with regional demand differences. In Fig. 4 A, the 2022 results indicate that the actual private pile capacity configuration was concentrated in core districts such as Yuzhong and Jiangbei, whereas peripheral districts such as Hechuan and Wanzhou received less than planned. This pattern suggests that construction first followed concentrated household demand in central areas but lagged behind in outlying counties. In contrast, the planning capacity configuration allocated private piles according to residential density and car ownership potential while also reserving room for future growth in peripheral areas.

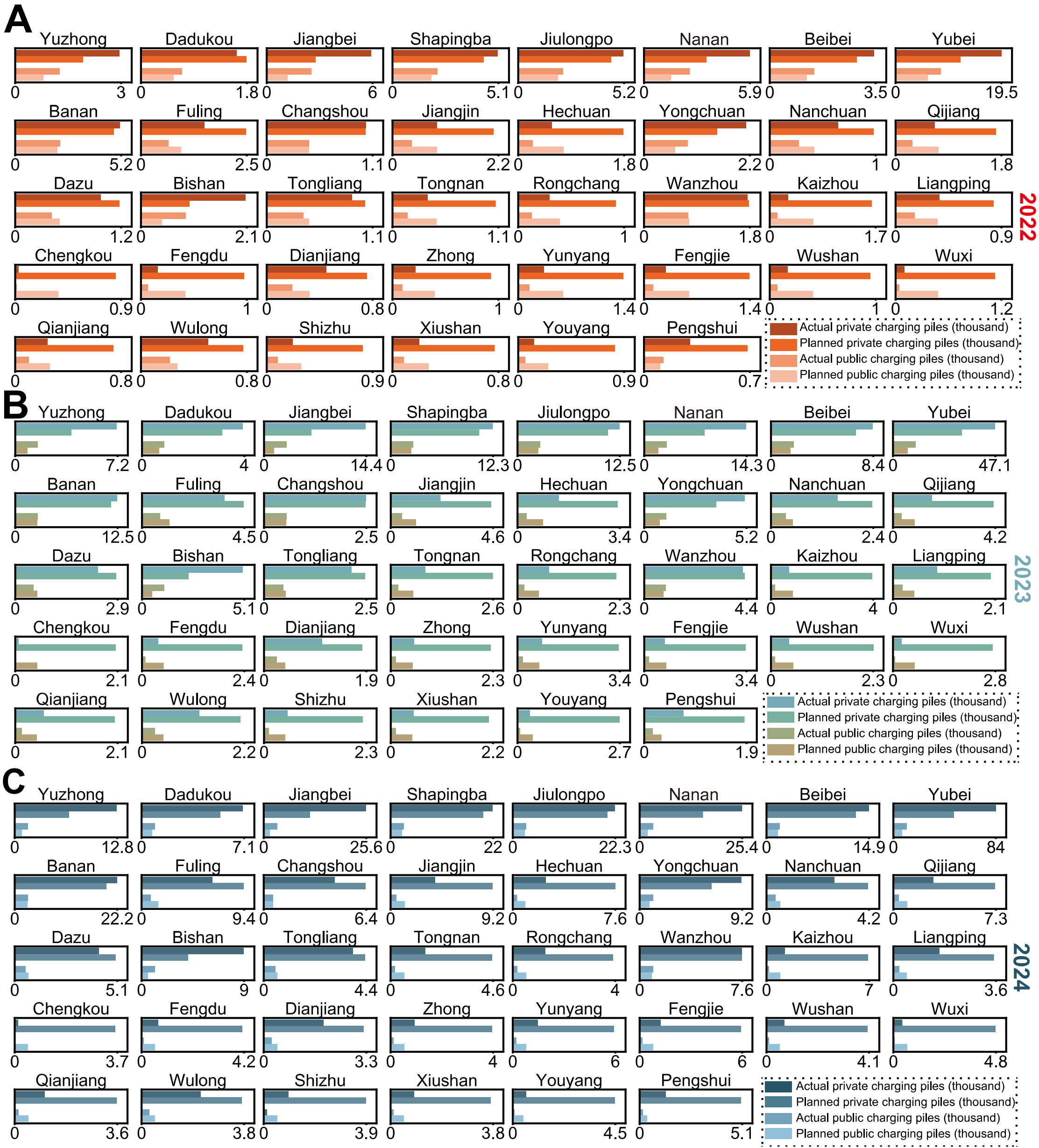

**Fig. 4.** Actual and planned capacity configurations of EV charging piles across districts and counties in Chongqing in (A) 2022, (B) 2023, and (C) 2024.

Fig. 4 B shows the 2023 charging capacity configuration. At the private-pile level, the actual capacity in many districts still deviated from the planned values. Several core districts remained below the planned level, while some near suburban districts slightly exceeded it, indicating an uneven construction pace that did not fully match the timing of demand release. For public piles, the actual capacity in all districts was far below plan, with shortages across both core and peripheral areas. This meant that high-frequency public travel needs, such as ride hailing and logistics in central areas, as well as basic public transport needs in outer counties, were insufficiently supported. The planning capacity configuration, however, differentiates public pile allocation by functional demand, increasing density in core areas and improving coverage in peripheral districts.

Fig. 4 C presents the 2024 comparison and reveals a similar structural imbalance in the actual capacity configuration. For private piles, some districts, especially Yubei, appeared to be oversupplied relative to reasonable demand, while many other core districts, including Dadukou and Jiangbei, remained below the level required by household charging needs. For public piles, the actual shortage was even more pronounced: with only a few exceptions, such as Chengkou and Shizhu, most districts did not reach the capacity needed for public service scenarios. In contrast, the planning capacity configuration aimed to correct these distortions by matching private piles to residential demand and public piles to travel intensity, thereby improving both fairness and efficiency in the spatial allocation of charging piles.

The comparison indicates that the actual configuration was reactive and spatially imbalanced, with construction delays in core areas, shortages in peripheral districts, and insufficient private and public charging piles. In contrast, the planning capacity configuration presented a more coordinated allocation framework with a forward-looking orientation. It determined district-level capacities more systematically and optimized the internal structure between private and public piles, ensuring better alignment with residential demand and public service functions. As a result, it provides a more rational and robust configuration for the sustainable development of charging piles. Taken together, these findings respond to Question 2 raised in Section 1.

### *3.3. Robustness and real-world validation of the planning framework*

This study developed and validated a charging capacity planning framework. The framework outperformed current real-world configurations in performance, spatial balance, and adaptability. However, a decision-support planning model should be evaluated not only by baseline performance but also by robustness to key parameter changes and by output accuracy and reproducibility. Fig. 5 presents a sensitivity analysis under key parameter changes. In Fig. 5 A, the CLTC served as the benchmark, with the Worldwide Harmonized Light Vehicles Test

Procedure (WLTP) and New European Driving Cycle (NEDC) for comparison. Across BEV, EREV, and PHEV representative models, changes in driving range standards altered estimated consumption levels but did not change the trend. For BYD Han, consumption rose from 4.5 GWh in late 2022 to 13.5 GWh in 2024 under the CLTC. Under the WLTP, it increased by 17.7% to 15.9 GWh, while under the NEDC, it decreased by 23.1% to 10.4 GWh, with the upward trend unchanged. Deepal SL03 showed the same pattern, rising from 5.5 GWh to 16.5 GWh under the CLTC, to 18.2 GWh under the WLTP and to 13.4 GWh under the NEDC, while preserving the same growth trend. The Changan UNI-V fluctuated more strongly, but its temporal pattern remained robust: under the CLTC, it peaked at approximately 2.7 GWh in 2023 and declined to approximately 1.8 GWh in 2024; under the WLTP, it rose to 3.3 GWh; and under the NEDC, it fell to 1.6 GWh, retaining the same rise-then-decline pattern. These results indicate that changes in driving range standards mainly affect magnitude rather than trend, supporting the robustness of the demand accounting model.

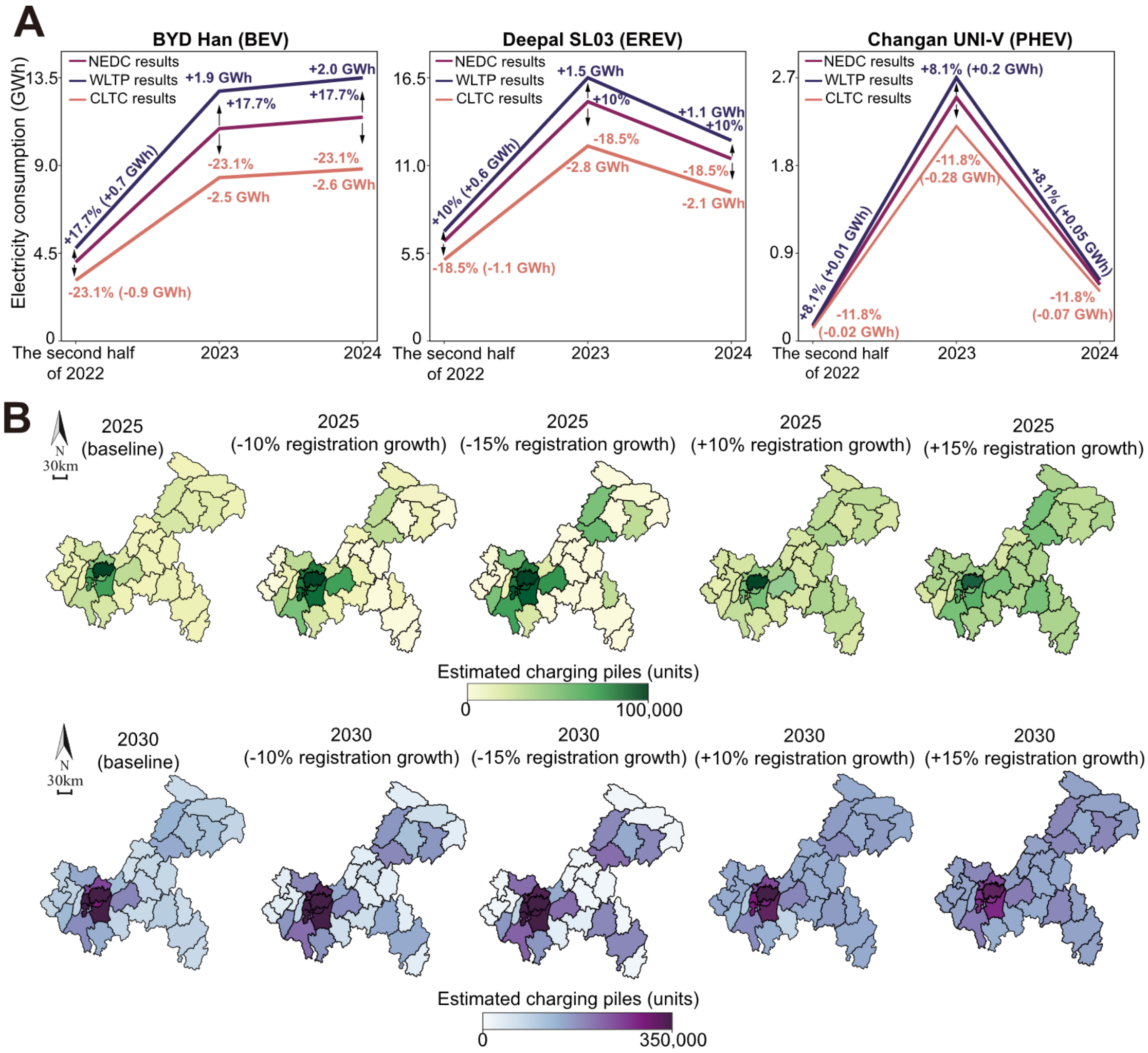

**Fig. 5.** Sensitivity analyses of EV electricity consumption and charging piles capacity configurations in Chongqing. (A) Sensitivity of electricity consumption for different EV types under NEDC, WLTP, and CLTC conditions, June 2022 to December 2024; (B) sensitivity of charging pile capacity configurations under varying EV registration growth rates in 2025 and 2030.

Fig. 5 B reports the sensitivity of charging pile capacity configuration to EV registration growth rates in 2025 and 2030. The spatial allocation pattern remained stable across scenarios, despite differences in absolute capacity. In the baseline case, central urban areas concentrated the highest charging capacity, while peripheral areas remained lower. When EV registration growth decreased by 10% or 15%, capacity became more concentrated in the core area, widening the central-peripheral gap and indicating increased spatial imbalance. In contrast, when the growth rate increased by 10% or 15%, capacity spread more evenly toward peripheral areas, and the distribution became less concentrated, although the central area still held the largest share. The same pattern was observed in 2030. A 10% or 15% decline strengthened

core dominance, whereas a 10% or 15% increase promoted a more balanced distribution. Even so, the relative hierarchy between central and peripheral areas did not change across scenarios. This shows that the model responded to demand scale changes through magnitude and balance adjustments, while its spatial distribution remained stable.

The detailed validation of the energy intensity estimation and the robustness of the HHO-based planning model are reported in Appendix E, where the estimated values closely match the observed data and the optimization results show stable convergence across repeated runs.

# 4. Discussion

This section summarizes the projected growth of the total number of charging piles in Chongqing from 2025 to 2030, the optimized spatial configuration of charging capacity across urban core, suburban, and county areas, and the corresponding policy implications.

## *4.1. Future pile capacity projections*

To forecast the future total number of charging piles in Chongqing and provide a data basis for subsequent capacity planning, this study developed a fractional-order grey model (FGM) [57] based on HHO optimization (HHO-FGM) to predict the total charging pile demand during the 15th Five-Year Plan period in Chongqing, and the principle of the HHO-FGM model is provided in Appendix F. Fig. 7 presents the results of the HHO-FGM model used to forecast the total number of charging piles in Chongqing from 2016 to 2030. This shows that the model achieved both stable parameter search and strong predictive performance for a small sample, nonlinear growth sequence. Fig. 7 A summarizes the optimization process for the model, showing that the objective function, measured by the validation set Mean Absolute Percentage Error (MAPE), dropped sharply in the early stage and then leveled off at approximately 0.02. This indicates that the HHO algorithm quickly identified a low error region and then refined the solution through local search. The fitted fractional order converged to approximately 0.1, which suggested that the model retained the advantages of the classic grey model while introducing only a mild fractional adjustment to better capture the growth pattern.

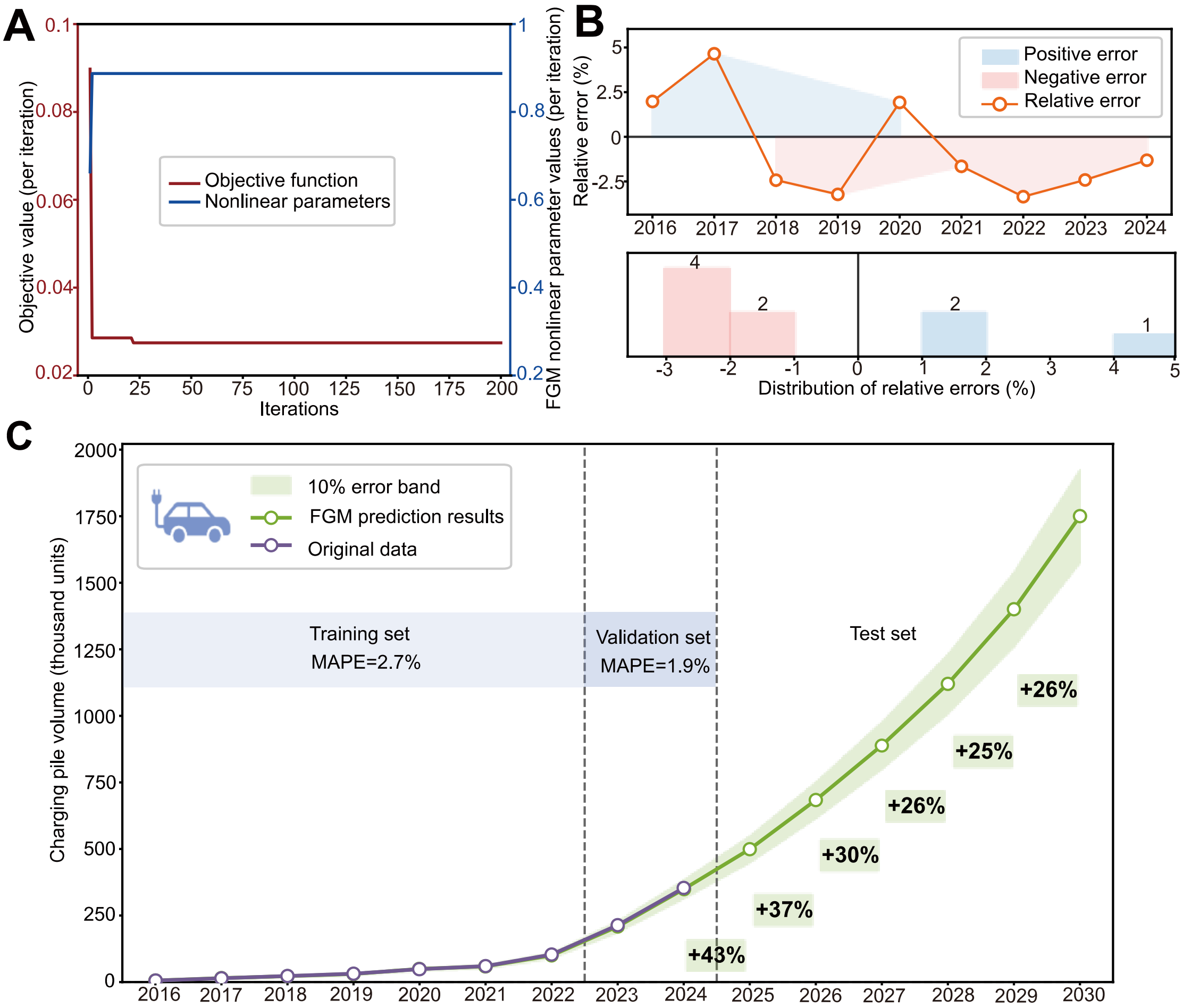

**Fig. 6.** Training performance of the HHO-FGM model and forecasts of the total number of charging piles in Chongqing from 2016 to 2030. (A) Convergence curve of the HHO-FGM model; (B) relative errors for in-sample data points and their distribution for the HHO-FGM model; (C) forecast results of the HHO-FGM model for the total number of charging piles.

Fig. 7 B reports the in-sample relative error and its distribution. The error was closely aligned with the development stages of the charging pile stock. In the early years, when the total number of charging piles was still very small and construction was unstable, the relative error reached a peak of approximately 4.5% in 2017. From 2018 to 2022, as growth became more stable, the error turned slightly negative and remained mostly within -3%. During 2023 and 2024, when expansion accelerated, the error became positive again but stayed within 2.5%. The distribution was concentrated in the minus 3% to 0 range, accounting for 66.7% of all years, while the remaining values were still confined to a narrow band. This pattern indicates that the model tracked the observed curve well and maintained high robustness across different growth stages.

Fig. 7 C shows the fitted historical series and the forecast path for 2025 to 2030. The model reproduced the observed data closely in the training period, with a MAPE of 2.7%, and performed even better in validation, where MAPE declined to 1.9%. For the forecast period, the annual growth rate was approximately 43% in 2025, reflecting the continuation of the rapid expansion seen in the previous stage. After that, growth gradually moderated to approximately 25 to 30% as the base became larger, which was consistent with the usual development pattern of charging pile deployment. By 2030, the total number of charging piles was projected to approach 1.8 million, suggesting that the forecast was not only statistically sound but also consistent with the broader policy and market logic of Chongqing's electrification transition.

### *4.2. Spatial distribution of optimized charging piles*

Based on the forecasts of energy demand and total charging piles capacity for 2025 to 2030, this study further distributed the total target across Chongqing's districts and counties to develop a sustainable charging piles capacity configuration that supports multiple objectives. Fig. 8 presents the charging pile capacity configurations from 2025 to 2030. In spatial terms, the total capacity was 499,000 in 2025 and was concentrated mainly in the central urban districts, while the surrounding counties remained relatively undersupplied. As the planning period advances, the total capacity would rise to 684,000 in 2026, 1.12 million in 2028, and more than 1.75 million by 2030. During this process, capacity in the central urban area continued to grow, while the surrounding counties also experienced gradual expansion. This pattern is consistent with the higher concentration of EVs and charging demand in the urban core, and it also reflects the outward spread of EV penetration. In this way, the charging pile capacity configurations help avoid a mismatch in which the core area becomes oversupplied while peripheral areas remain insufficiently served.

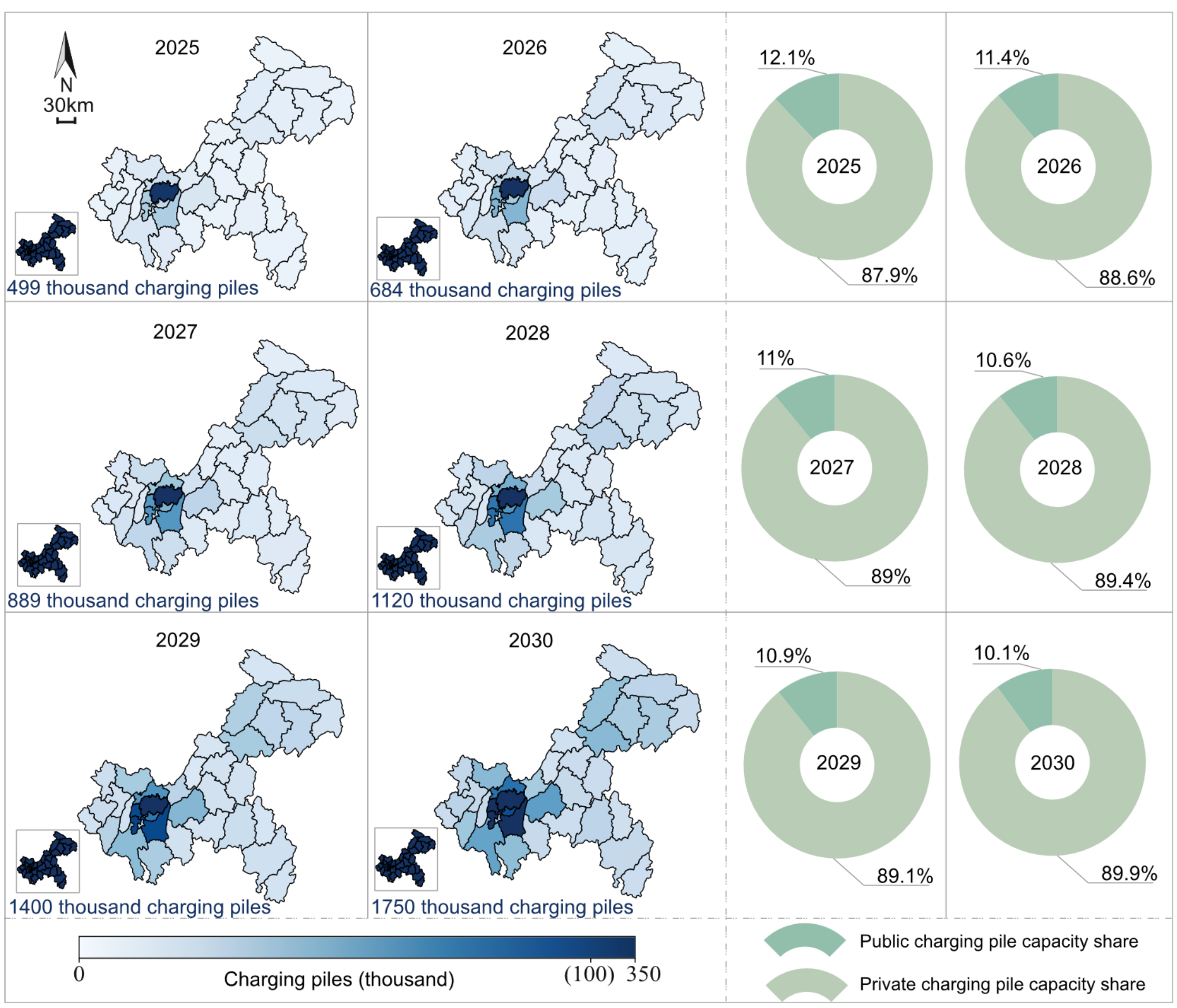

**Fig. 7.** Charging pile capacity configurations in Chongqing from 2025 to 2030.

From a structural perspective, the share of public charging piles declined slightly from 12.1% in 2025 to 10.1% in 2030, whereas the share of private charging piles increased from 87.9% to 89.9%. This did not indicate a reduction in public charging capacity. Rather, private charging expanded more rapidly. Specifically, public charging capacity increased from approximately 60,400 in 2025 to approximately 176,800 in 2030, while private charging capacity grew from 438,600 to 1.57 million. The faster growth of private charging corresponded to the expansion of privately owned EVs, while the steady increase in public charging supported charging demand in public and commercial settings. Overall, the charging-pile capacity configuration maintains a stable balance between private and public charging, keeps the share of public charging above 10% throughout the study period, and strengthens the alignment between charging infrastructure supply and long-term demand evolution.

Fig. G1 in Appendix G illustrates how population size and land area jointly shape the allocation of charging piles across Chongqing's districts and counties from 2025 to 2030. In Fig.

G1 A, these two factors jointly determine public charging-pile capacity through both demand intensity and service coverage. Population size primarily drives demand levels, as densely populated areas support more frequent public travel, stronger commercial activity, and heavier utilization by taxis, ride-hailing vehicles, and logistics fleets. As a result, districts such as Liangjiang maintain high public charging-pile capacity throughout the study period. In contrast, counties with smaller populations, such as Pengshui and Youyang, maintain lower capacities because public charging demand is less concentrated. Land area mainly affects the required service radius. Larger districts such as Wanzhou and Kaizhou require more spatially distributed charging coverage, leading to higher public charging-pile capacity than compact districts with similar population levels. Over time, population size plays a stronger role during the earlier years, whereas the influence of land area becomes more pronounced in the later stages as expanding suburban areas require broader spatial coverage.

Fig. G1 B shows a similar but more refined pattern for private charging piles. Here, population size mainly determines the overall demand base and residential installation density, whereas land area affects how private charging facilities adapt to different housing forms and spatial conditions. In highly populated districts such as Liangjiang, private charging-pile deployment evolves from basic community-level coverage in 2025 to much denser household-level support by 2030, particularly in large residential communities. In sparsely populated counties such as Pengshui, private charging piles remain centered on community-shared supply, thereby avoiding unnecessary duplication. Land area further shapes the spatial mode of provision. Larger districts such as Wanzhou rely on a mixed strategy that combines reserved charging spaces in new housing developments, additional shared charging piles in older communities, and centralized charging hubs in township settlements. Smaller urban-core districts such as Yuzhong make greater use of vertical space and shared facilities to increase charging supply within limited residential land. Overall, Fig. G1 B indicates that private charging capacity evolves from simple scale matching toward more precise spatial adaptation, with population size guiding demand growth and land area shaping facility form and spatial layout.

Based on the projected electricity demand and total charging-pile capacity from 2025 to 2030, this study further allocated charging capacity across Chongqing's districts and counties to establish a sustainable configuration that balances scale, structure, and spatial equity. The results show a steady increase in total charging capacity, with the urban core remaining the primary supply center while surrounding counties gradually receive greater support. This pattern helps avoid both overconcentration in central districts and underprovision in peripheral areas. At the same time, private charging grows faster than public charging, while the share of public charging remains above 10% throughout the planning period, indicating a stable balance between household and shared charging demand. Overall, population size and land area jointly shape the spatial allocation of public and private charging piles, leading to a more refined and adaptive planning pattern. The findings presented in this section respond to Question 3 raised in Section 1.

### *4.3. Policy implications*

After the completion of theoretical construction, empirical analysis, and uncertainty validation of charging piles capacity planning, policy guidance and institutional support remain essential for implementation. At the national level, China has gradually formed a policy system covering planning guidance, financial incentives, technical standards, and electricity pricing. Chongqing's policy framework has also evolved from initial exploratory measures from 2015 to 2020 to accelerated deployment from 2021 to 2023 and then to a stage of high-quality development since 2024, with a clear shift from scale expansion to quality improvement, especially through ultrafast charging initiatives. Overall, the policy orientation has become more explicit, the technical path more distinct, and the policy instruments more diversified. However, several bottlenecks still constrain implementation. Demand forecasting remains overly dependent on simplified parameter assumptions and annual scale estimates, which cannot fully capture vehicle type differences or seasonal fluctuations, such as the approximately 30% winter increase in charging demand. Capacity allocation also lacks a robust multi-objective balancing mechanism, which helps explain the persistent mismatch between overconcentration in core urban areas and lagging provision in peripheral districts. In addition, land, electricity, and capital resources are not well coordinated, cross-department planning and data sharing are weak, approval procedures are fragmented, and market incentives remain insufficient, leaving many operators with low utilization and weak profitability.

Based on these findings, the planning method should shift from estimation grounded in experience toward forecasting that is driven by data and dynamically adjusted. First, the data foundation needs to be strengthened by integrating vehicle registration records, traffic survey information, and grid load monitoring data so that demand estimation can be differentiated by vehicle type and periodically recalibrated. The empirical results in this study show clear differences in energy intensity among the BEV, EREV, and PHEV models, with the highest values reaching approximately 37 kWh/100 km, which confirms that a single fixed coefficient is no longer adequate. Second, capacity configuration should be evaluated through multiple criteria, including energy intensity, grid carrying capacity, service coverage, and regional fairness [58], rather than relying only on configuration based on administrative divisions. Third, planning should adopt a rolling adjustment mechanism that combines annual monitoring, midterm evaluation, and subsequent optimization because demand growth, technology change, and seasonal variation all create meaningful deviations from static plans.

The spatial strategy for implementation should also become more differentiated. In central urban areas, the priority should be stock optimization rather than simple expansion, with greater emphasis on shared charging in old residential communities, strict enforcement of charging facility requirements in new housing projects, and targeted deployment of public fast charging near commercial centers and transport hubs. In new suburban districts, charging piles should move in step with urban expansion, with the advance reservation of charging space in land use planning and stronger support for fast charging deployed along transport routes. In county areas, the main goal should be basic coverage and gradual supplementation, focusing on county centers, key towns, and transport nodes while avoiding overinvestment in areas with weak demand. At the same time, coordination between vehicles, charging piles, and the power grid must be strengthened through joint planning, unified approval, and data sharing. Grid capacity information should be disclosed more regularly, and flexible solutions such as smart charging and integrated solar storage charging systems should be encouraged where conditions are suitable.

Policy incentives and institutional safeguards should be improved in parallel. Financial support should move from one-time construction subsidies toward a full chain framework covering construction, operation, and service quality, with differentiated support for central city renovation, suburban expansion, and county-level basic coverage. Operational incentives should be tied to utilization, uptime, and repair efficiency so that public funds reward actual service performance rather than only installed capacity. Price signals should be refined through better peak and off-peak pricing, which can encourage users to charge at lower load periods and improve grid balance. Finally, binding standards, especially for new project charging provisions, ultrafast charging interoperability, safety management, and data interfaces, should be enforced more strictly. Public participation and monitoring should also be institutionalized through open information platforms, regular evaluation of supply demand matching, and feedback mechanisms that support continual policy correction and development that is characterized by high quality.

## 5. Conclusion

This study developed a framework for city-scale EV charging assessment and public–private charging pile planning in Chongqing. A bottom-up approach using vehicle registration, technical, and climatic data was applied to estimate historical charging demand. The estimated demand was incorporated into a multi-objective optimization model for public and private charging pile planning, with the HHO algorithm balancing cost efficiency, grid stability, and service adequacy. The analysis covers 2022–2030 and supports the evaluation of current deployment and future capacity needs. The key findings of this study are listed below.

### *5.1. Key findings*

- **From June 2022 to December 2024, Chongqing's total monthly EV electricity demand surged from 18.9 GWh to 57.5 GWh, representing a 204% increase and highlighting rapid market penetration.** This growth exhibited pronounced seasonal volatility, with consumption peaking in winter months due to increased heating demand and reduced battery efficiency in cold weather. While BEVs remained the primary consumers of electricity, their share of total monthly consumption declined from 63.4% in late 2022 to 42.3% by the end of 2024, as the combined share of PHEVs and EREVs expanded to 57.6%. Analysis that focused on specific models revealed that models with high registrations, such as the Tesla Model Y and BYD Song PLUS, dominated regional demand, especially Model Y, which showed the highest volatility that reached 8.3 times the consumption between its maximum and minimum months. This shift toward diversity in powertrains necessitates a transition from charging pile planning focused on vehicles to planning for specific technologies to accommodate varying electricity intensities.
- **A comparative evaluation of historical charging pile deployment from 2022 to 2024 revealed significant spatial imbalances, where actual construction was heavily concentrated in the urban core at the expense of peripheral service coverage.** In 2024, the actual configuration showed a marked structural mismatch: Yubei appeared oversupplied relative to residential demand, while other core districts such as Dadukou and Jiangbei remained below the levels required for household charging. Furthermore, public charging capacity across nearly all districts consistently lagged behind planned levels, failing to adequately support public travel and logistics with high frequency. The optimized configuration proposed in this study achieved a superior comprehensive performance score of 0.28 in 2024, which stood in contrast to the 0.65 of the actual deployment, by effectively balancing efficiency, regional fairness, and grid stability as demand expanded.

- **Between 2025 and 2030, Chongqing's charging piles were projected to enter a sustained growth phase, with the total number of piles reaching approximately 1.8 million units by 2030.** The optimized 2030 planning framework maintained a stable 9:1 private-to-public charging ratio, with private piles expanding to 1.57 million to meet residential needs and public capacity reaching 176,800 units to support commercial settings. This synergetic deployment was expected to significantly mitigate urban–rural service disparities and improve system-wide resilience, delivering a 24.8% reduction in grid load fluctuations and a 6.6% improvement in energy supply potential relative to the 2024 baseline. These results underscore that a multi-objective approach with rigid regional equity constraints can achieve a more layered, balanced, and grid-friendly spatial structure across the "Core-Suburban-Exurban" hierarchy.

### *5.2. Future work*

Building on the present framework for city-level charging pile planning, future studies may further advance in two directions. First, the analytical design can be strengthened by incorporating richer and more dynamic data sources, including real-time traffic flows, travel chain information, individual charging preferences, vehicle connectivity data, mobile signaling data, and charging application records, to develop a more granular human-vehicle-charger interaction model with higher spatial and temporal resolution. This would provide a more accurate basis for capturing the microlevel mechanisms and uncertainty of charging demand while also allowing future studies to examine long-term factors such as battery degradation and the evolution of charging technologies. Second, the proposed approach can be extended to cities with different spatial structures, power grid conditions, and policy environments across eastern, central, and western China. Comparative studies across multiple urban cases would help assess the adaptability of the framework, refine its implementation under diverse local conditions, and summarize more general planning patterns and policy tools, thereby offering broader support for the high-quality development of charging infrastructure at the national level.

## Appendix

Please find the appendix in the supplementary materials (e-component).

## Acknowledgments

None.

## Declaration of interests

The authors declare that they have no competing interests.